\documentclass[usenatbib,fleqn,useAMS]{mnras}

\pdfoutput=1
\usepackage{hyperref,graphicx,natbib,times,amssymb,amsmath,pstricks,soul,color, epstopdf}
\usepackage{aas_macros}
\usepackage{bm,url}
\usepackage{textcomp}

\usepackage{cleveref}
\usepackage[T1]{fontenc}
\usepackage{ae,aecompl}
\usepackage{mathtools}

\newcommand  *{\B}[1] {\boldsymbol{#1}}

\title[Efficient dust ring formation in misaligned circumbinary discs]{Efficient dust ring formation in misaligned circumbinary discs}
\author[Hossam Aly \& Giuseppe Lodato]
       {Hossam Aly$^{1,2}$\thanks{Email: hossam.aly@gmail.com}
        and Giuseppe Lodato$^1$ \vspace{0.05in}\\ 
        $^1$ Universit\`a degli Studi di Milano, Via Giovanni Celoria 16, 20133 Milano, Italy\\
        $^{2}$ Univ Lyon, Univ Claude Bernard Lyon 1, Ens de Lyon, CNRS, Centre de Recherche Astrophysique de Lyon UMR5574, F-69230, Saint-Genis-Laval, France}
\date{Accepted  2019 December 14.
      Received 2019 December 13;
      In original form 2019 August 24;
      }
\pagerange{\pageref{firstpage}--\pageref{lastpage}}
\pubyear{2019}
\begin{document}
\label{firstpage}
\pagerange{\pageref{firstpage}--\pageref{lastpage}}
\maketitle
% Abstract of the paper
\begin{abstract}
Binary systems exert a gravitational torque on misaligned discs orbiting them, causing differential precession which may produce disc warping and tearing. While this is well understood for gas-only discs, misaligned cirumbinary discs of gas and dust have not been thoroughly investigated. We perform SPH simulations of misaligned gas and dust discs around binaries to investigate the different evolution of these two components. We choose two different disc aspect ratios: A thin case for which the gas disc always breaks, and a thick one where a smooth warp develops throughout the disc. For each case, we run simulations of five different dust species with different degrees of coupling with the gas component, varying in Stokes number from 0.002 (strongly coupled dust) to 1000 (effectively decoupled dust). We report two new phenomena: First, large dust grains in thick discs pile up at the warp location, forming narrow dust rings, due to a difference in precession between the gas and dust components. These pile ups do not form at gas pressure maxima, and hence are different from conventional dust traps. This effect is most evident for St $\sim$ 10 - 100. Second, thin discs tear and break only in the gas, while  dust particles with St $\geq$ 10 form a dense dust trap due to the steep pressure gradient caused by the break in the gas. We find that dust with St $\leq$ 0.02 closely follow the gas particles, for both thin and thick discs, with radial drift becoming noticeable only for the largest grains in this range.
\end{abstract}

%%%%%%%%%%%%%%%%%%%%%%%%%%%%%%%%%%%%%%%%%%%%%%%%%%
%%%%%%%%%%%%%%%%%%%%%%%%%%%%%%%%%%%%%%%%%%%%%%%%%%

%%%%%%%%%%%%%%%%% BODY OF PAPER %%%%%%%%%%%%%%%%%%

\section{Introduction}
Stars form as a result of molecular cloud cores collapsing under the effects of self-gravity. Many of them form in binary or multiple systems. Conservation of angular momentum predicts the formation of flattened discs as the molecular cloud core collapses. These discs of gas and dust are thought to be the location where planet formation takes place. For close binary systems, we expect circumbinary discs to form. Recent observations of circumbinary planets confirm these expectations \citep{DoyleEtal2011,WelshEtal2012,OroszEtal2012a,OroszEtal2012b,WelshEtal2015,KostovEtal2013,KostovEtal2016}.

Due to the random nature of star formation \citep{Bonnell&Bastien1992,OffnerEtal2010,Bate2010,Bate2018}, at least some circumbinary discs are expected to form misaligned to the binary orbital plane. These expectations have been confirmed by recent observations; for example GG Tau A \citep{Khoeler2011,AndrewsEtal2014,AlyEtal2018}, KH 15D \citep{Chiang&Murray2004,WinnEtal2004,Lodato&Facchini2013,SmallwoodEtal2019,Fang2019}, IRS 43 \citep{BrinchEtal2016}, and L1551 NE \citep{TakakuwaEtal2017}. Stellar binaries exert gravitational torques on inclined circumbinary discs causing radially differential precession. In the presence of dissipation, differential precession will cause the evolution of warps within the disc which can either diffuse or propagate in a wave-like manner depending on disc thickness and viscosity \citep{Papaloizou&Pringle1983,Papaloizou&Lin1995,Ogilivie1999,LodatoPringle2007,LodatoPrice2010}. Dissipation and warp evolution lead to planar alignment of the disc with respect to the binary orbital plane, either in a prograde or retrograde sense depending on the initial inclination \citep{KingEtal2005,NixonEtal2011}, with a possibility of tearing the disc whenever the gravitational torque exceeds the viscous torque \citep{NixonEtal2013,DoganEtal2018}. For inclined low mass discs around eccentric binaries, polar alignment is possible \citep{Farago&Laskar2010,AlyEtal2015,Martin&Lubow2017,Zanazzi&Lai2018}, a configuration that has been recently observed in the HD98800 system \citep{Kennedy2019}.

While recent studies have explored the evolution of inclined gas discs around binaries, the interplay between gas and dust in such an environment is yet to be fully explored. The standard theory for planet formation is the well known `core accretion' theory \citep{Safronov1969,Goldreich&Ward1973}. In this context, a metal rich core forms and grows through collision of planetesimals up to a point where it reaches a high enough mass for runaway gas accretion to take place. In this scenario, dust grains must grow through coagulation and clumping from micron sizes up to kilometer size planetesimals \citep{TestiEtal2014}. However, this growth is thought to be hampered by the radial drift of dust particles caused by drag forces \citep{Whipple1972,AdachiEtal1976,Weidenschilling1977}. This radial drift of dust particles is caused by the the velocity difference between the gas and dust components: while dust particles orbit at Keplerian speed, gas experiences an outward pressure gradient causing it to orbit at sub-Keplerian speed. This velocity difference causes a head wind on the dust particles causing them to drift inwards. This is the so called `meter sized barrier' problem to planet formation, although the naming came about as this was first applied to a Minimum Mass Solar Nebula (MMSN), for which the critical size dust particles are close 1 meter. A more general way of identifying critical dust sizes is by making use of the Stokes number St, as the ratio of the stopping time to the orbital time:
\begin{equation}
    \mathrm{St}=\Omega_k \tau_s
\end{equation}
Where $\Omega_k$ is the keplerian angular speed and $\tau_s$ is the characteristic stopping time, which is a measure of the coupling due to the drag force. For typical protoplanetary discs and particle sizes less than 1 meter, The gas mean free path is much larger than the dust particle size ($\lambda_{g}/s\leq 4/9$, where $\lambda_{g}$ is the gas mean free path and $s$ is the dust particle size) and the drag force is well described by the Epstein regime \citep{Epstein1924}. In this regime the stopping time can be described by
\begin{equation}
   \tau_s = \frac{\rho_{d} s}{\rho_g v_{th}}
\end{equation}
where $\rho_d$ is the dust intrinsic density, $\rho_g$ is the gas density,  $v_{th} = \sqrt{8/\pi} c_s$ id the mean thermal velocity and $c_s$ is the sound speed. The Stokes number then becomes
\begin{equation}
    \mathrm{St}=\frac{\pi}{2}\frac{\rho_d s}{\Sigma_g}
\end{equation}
with $\Sigma_g$ denoting the gas surface density.
Particles with St $\sim 1$ experience the strongest radial drift \citep{BrauerEtal2007}. This has severe implications on planet formation; namely, mm-cm sized grains drift inwards within a small fraction of the disc lifetime before planetesimals can form. 
%Moreover, recent (sub-)mm observations of evolved protoplanetary discs show that this is not the case \citep{TestiEtal2014}. 

A proposed solution to the radial drift barrier to planet formation is that dust particles get trapped in gas local pressure maxima \citep{NakagawaEtal1986}. Indeed the drift velocity vanishes exactly when the pressure gradient is zero. A few mechanisms were suggested to form these dust traps, such as vortices \citep{Barge&Sommeria1995,Klahr&Henning1997}, gaps forged by planets \citep{Paardekooper&Mellema2004,PinillaEtal2012}, spirals induced by gravitational instabilities \citep{Rice2004,Rice2006} and self induced dust traps when taking into account the dust backreaction on the gas as well as dust growth and fragmentation \citep{GonzalezEtal2017}.

In this paper, we study the interaction between gas and dust in inclined circumbinary discs. We investigate the effects of the resulting precession and induced warps on the dust evolution and whether this could offer a solution to the radial drift problem. 
We perform Smoothed Particle Hydrodynamics (SPH) simulations of misaligned gas and dust discs around an equal mass circular binary, varying the dust St and disc aspect ratio. We present the details of the numerical simulations in Section~\ref{sec:setup}. In Section~\ref{sec:large_grains} we show and discuss our results for large dust particles and discuss how these can be connected with observations in Section~\ref{sec:discuss}. We then present our results for small dust grains in Section~\ref{sec:small_grains} before we conclude our findings in Section~\ref{sec:conclusion}.
%%%%%%%%%%%%%%%%%%%%%%%%%%%%%%%%%%%%%%%%%%%%%%%%%%
%%%%%%%%%%%%%%%%%%%%%%%%%%%%%%%%%%%%%%%%%%%%%%%%%%
\section{Hydrodynamic Simulations}
\label{sec:setup}
We perform a suite of numerical simulations using the SPH code PHANTOM \citep{PriceEtal2018}. PHANTOM implements two different ways for treating gas-dust mixtures: a two-fluid model for large ($\mathrm{St} \geq 1$) dust particles and a one-fluid model for small ($\mathrm{St} < 1$) dust grains. In the two-fluid model \citep{Laibe&Price2012}, the two species are treated with two different sets of particles which carry information about the properties of their corresponding type. The usual SPH density summation is carried out over neighbours of the same particle type. Two sets of governing equations are solved and a drag term is added to account for the interaction between gas and dust. The drag term uses a `double hump' kernel, in contrast to the usual SPH kernel implementation (details of the algorithm can be found in \citet{PriceEtal2018} and \citet{Laibe&Price2012}). The algorithm computes the Knudsen number 
\begin{equation}
    \mathrm{Kn}=\frac{9\lambda_g}{4s}
\end{equation}
where $\lambda_g$ is the gas mean free path and $s$ is the grain size. A drag regime is selected automatically based on the particle Kn; Epstein drag \citep{Kwok1975} is computed for particles with $\mathrm{Kn} \geq 1$ and Stokes drag is computed for $\mathrm{Kn} < 1$, with a smooth transition between the two regimes. Importantly, the code calculates the stopping time -- a measure of the decay of the differential velocity between the two phases due to the drag force -- between a dust--gas particles pair according to:
\begin{equation}
    \mathrm{t_{stop}}=\frac{\rho_{g}\rho_{d}}{K(\rho_{g}+\rho_{d})}
\end{equation}
Where $K$ is a drag coefficient depending on the drag regime. The timestep is governed by the stopping time between two particles, making the two-fluid model extremely computationally expensive for low St simulations.

In the one-fluid model, on the other hand, 
 both dust and gas are treated as one mixture with the same set of governing equations, with an evolution equation for the dust fraction. In this method, the timestep is governed by the inverse of the stopping time, which becomes computationally expensive in the large dust particle regime, where the two-fluid model is more appropriate \citep{Price&Laibe2015}. This algorithm has been used and verified extensively in previous studies, for example in \citet{DipierroEtal2015,TriccoEtal2017,RagusaEtal2017,NealonEtal2019}. 

We perform simulations for 5 dust species spaning a wide range of St. For each dust species, we perform 2 simulations with a thick $H/R=$ 0.1 and thin $H/R=$ 0.05 disc aspect ratio at the inner radius. In all our simulations, We use a disc viscosity coefficient \citep{ShakuraSunyaev1973} $\alpha=0.01$. The binary stars are modeled using equal mass sink particles with a total mass of 1 M$_\odot$ and accretion radius of $r_{\rm acc}=0.5a$, where $a$ is the binary separation. The discs extend from $R_{\rm in}=1.5a$ to $R_{\rm out}=15a$ and are initially flat with an inclination of $\Theta=30$ and a total (gas + dust) mass of 0.02 M$_\odot$, with a dust to gas ratio 0f $0.01$. All our simulations are safely with the gravitational stable regime, therefore we do not employ self gravity in any of these computations. 

In all our simulations we use a gas density radial profile as a power law with an index $p=-1.5$. We assume a locally isothermal disc, where the sound speed $c_s$ varies only with disc radius as power law with an index of $q=-0.75$. This choice of parameters ensures that the disc is uniformly resolved at least initially \citep{LodatoPrice2010} and that the resulting physical viscosity coefficient is constant with radius. The sound speed power law also implies an aspect ratio radial profile $H/R \propto R^{-1/4}$. We use 1 million SPH gas particles, ensuring that the disc thickness is resolved by several smoothing length at any radius. For the high St cases modeled with the 2-fluid approach, we represent the dust with 300,000 particles.

The thick disc cases as well as all the cases with the 1-fluid model ran for more than 900 binary orbits ($\sim 3\times 10^4$ years). Cases with thin discs underwent disc breaking and tearing which caused a substantial decrease in the time step and hence more computational wall time. Nevertheless most of these simulations ran for more 500 binary orbits. We summarize all the simulations parameters as well as the final time in Table~\ref{table:parameters}.

It will be useful to define the two angles defining a warped disc, that is the tilt and twist angles. If we define as $\hat{\bf l}$ the unit vector corresponding to the disc angular momentum (defining the direction perpendicular to the local disc plane), we have that (e.g., see \citealt{FacchiniEtal2013}):
\begin{equation}
    \hat{\bf l} = (\cos\beta\sin\Theta,\sin\beta\sin\Theta,\cos\Theta),
\end{equation}
where $\Theta$ is the tilt angle and $\beta$ is the twist angle. A disc is warped whenever $\Theta$ varies with radius, and is twisted whenever additionally $\beta$ varies with radius.

\begin{table}
\begin{tabular}{ |c|c|c|c| } 
 \hline
  $H/R$ & Approx St & Dust Model & End Time (in binary orbits) \\ 
 \hline
 0.1 & 10 & 2-fluid & 900 \\ 
 0.1 & 100 & 2-fluid & 900 \\ 
 0.1 & 1000 & 2-fluid & 870 \\
 0.1 & 0.02 & 1-fluid & 1000 \\
 0.1 & 0.002 & 1-fluid & 1000 \\
 0.05 & 10 & 2-fluid & 150 \\
 0.05 & 100 & 2-fluid & 500 \\ 
 0.05 & 1000 & 2-fluid & 500 \\
 0.05 & 0.02 & 1-fluid & 1000 \\
 0.05 & 0.002 & 1-fluid & 1000 \\
 \hline
\end{tabular}
\caption{Parameters set for the simulations suite}
\label{table:parameters}
\end{table}

%%%%%%%%%%%%%%%%%%%%%%%%%%%%%%%%%%%%%%%%%%%%%%%%%%
%%%%%%%%%%%%%%%%%%%%%%%%%%%%%%%%%%%%%%%%%%%%%%%%%%
\section{Large Dust Grains}
\label{sec:large_grains}
In this Section we first present the results obtained with weakly coupled dust grains, with $St>1$, in the two gas configurations of a thick ($H/R=0.1$) and thin ($H/R=0.05$) disc. Strongly coupled dust simulations are presented later in Section \ref{sec:small_grains}.
%%%%%%%%%%%%%%%%%%%%%%%%%%%%%%%%%%%%%%%%%%%%%%%%%%
\subsection{Thick discs}
\label{sec:thick}
\begin{figure*}
  \begin{center}
    \resizebox{170.0mm}{!}{\mbox{\includegraphics[angle=0]{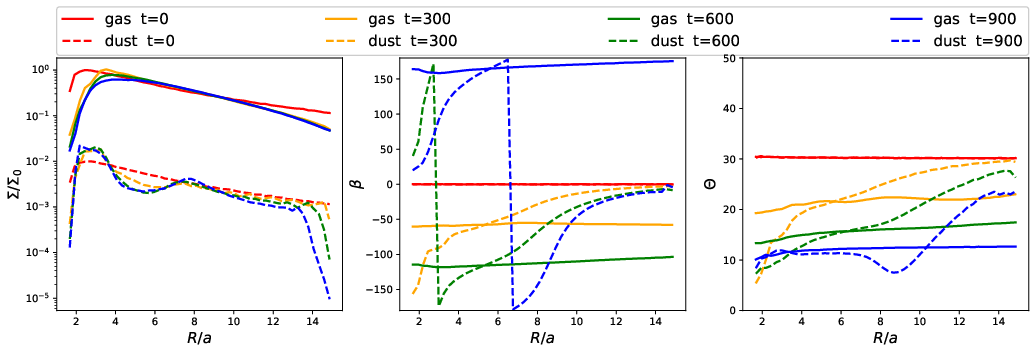}}}
    \resizebox{170.0mm}{!}{\mbox{\includegraphics[angle=0]{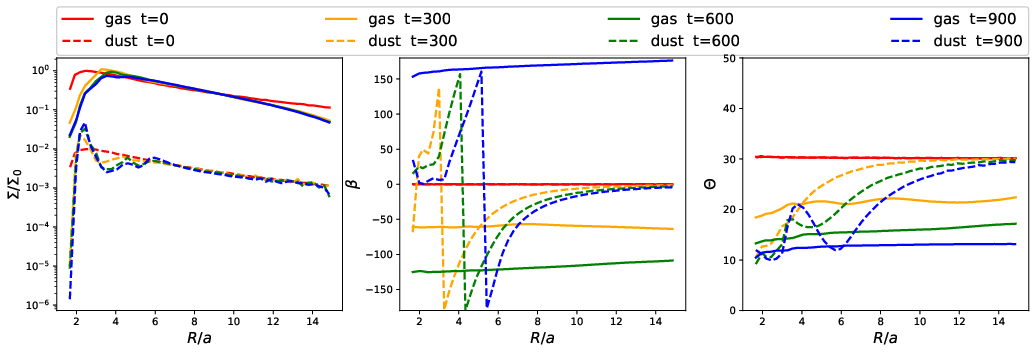}}}
    \resizebox{170.0mm}{!}{\mbox{\includegraphics[angle=0]{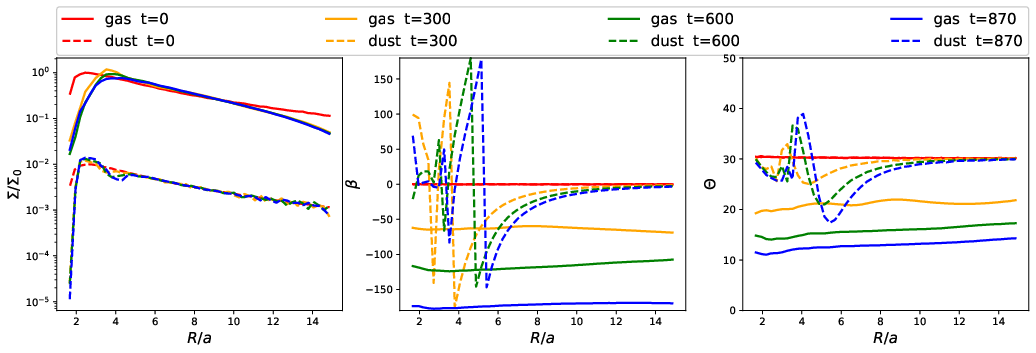}}} 
 \caption{\label{fig:largethick}
 Radial profiles of (left to right) surface density, twist, and tilt angles for gas (solid) and dust (dashed) at 4 different snapshots for $H/R=$ 0.1 cases. Top to bottom: $St \sim 10$, 100, and 1000.}
 \end{center}
\end{figure*}
\begin{figure}
\resizebox{85.0mm}{!}{\mbox{\includegraphics[angle=0]{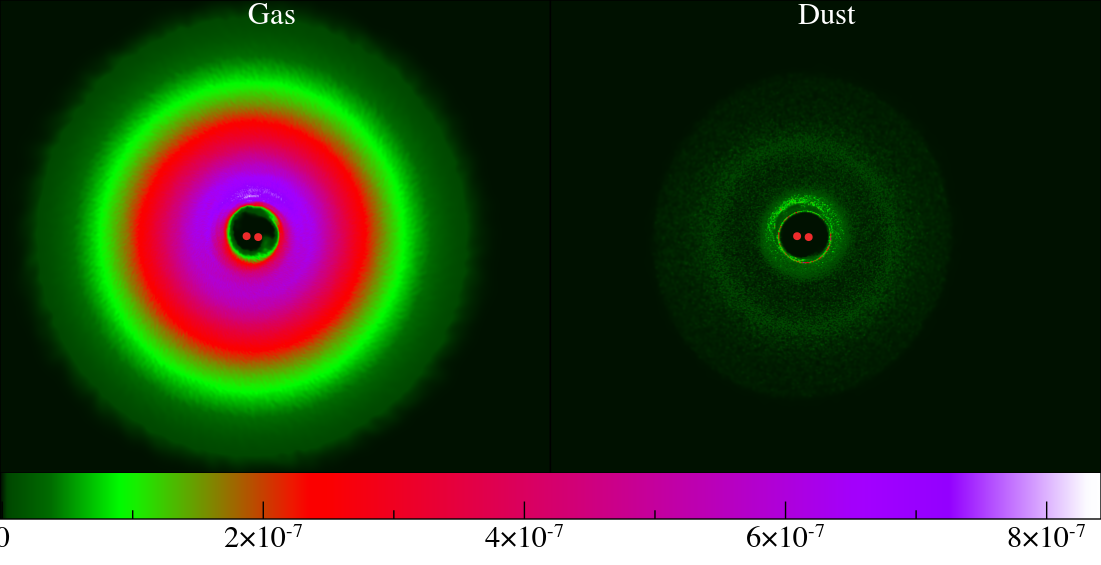}}}
\resizebox{85.0mm}{!}{\mbox{\includegraphics[angle=0]{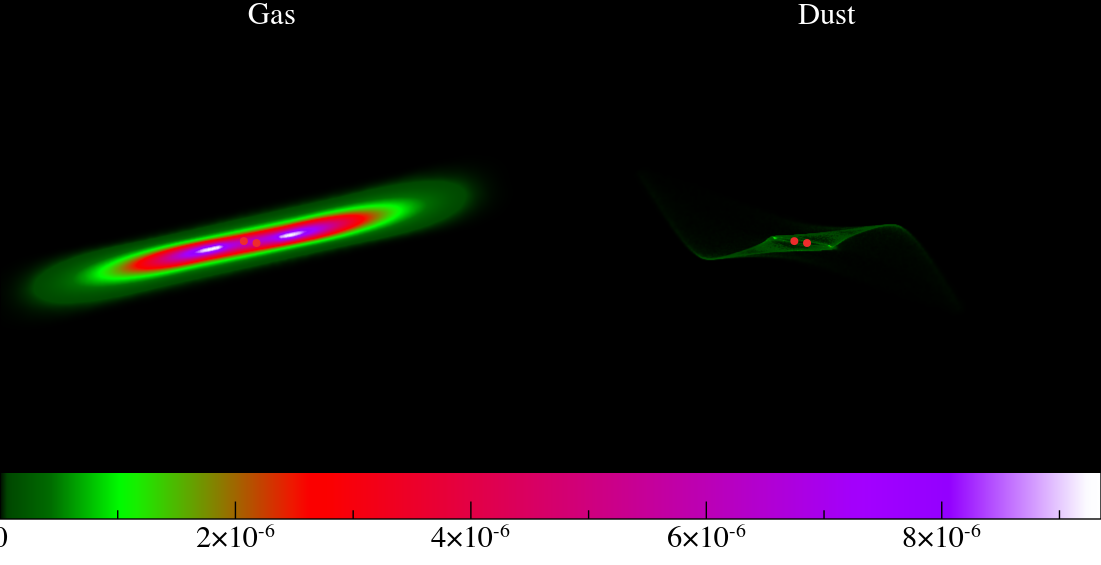}}}
\caption{\label{fig:dens_10cmthick}
 Density columns for gas (left panel) and dust (right panel) at the final snapshot for the St $\sim 10$ dust and $H/R=$ 0.1 case. Top: $xy$-plane. Bottom: $xz$-plane}
\end{figure}

We present the surface density, twist, and tilt radial profiles at 4 different times for the $H/R=$ 0.1 and $St \sim 10$, 100, and  1000 cases (top to bottom) in Fig.~\ref{fig:largethick}. Starting with the top row (St = $10$), the surface density profile shows a dust pile up in the inner disc at $R\sim 3a$, with another less pronounced bump at $R\sim 8a$. We note that neither of these dust pile ups occur at gas density peaks, implying that the underlying mechanism is not the usual trapping at pressure maxima. The middle panel shows the twist angle radial profile for the same 4 snapshots. Here we can clearly see that the gas and dust precess at different rates. The gas disc, which communicates viscously with the outer disc, tends to precess almost rigidly with only a few degrees difference in the twist angle between the inner and outer parts of the disc. The dust on the other hand does not communicate viscously, and can only experience the binary gravitational torque cause the differential precession, as well as the drag force from the gas component. Hence the dust displays a greater radial gradient in precession rate, with the inner part leading the gas disc and the outer part lagging behind the gas. The precession profile shows the gas and dust are out of phase for most of the radial extent of the disc. We can see that the intersection of the twist profiles occurs at $R\sim 5.5a$. This coincides with a dip in the dust surface density profile shown in the upper left panel. This means that the intersection between the gas and dust discs produces a head wind on the dust particles that causes radial drift at this radius. However, when dust particles migrate inwards, they go out of phase with the gas and pile up because of the reduced drag force. The right panel shows the tilt angle radial profile for the gas and dust. We see that the gas smoothly aligns with the binary plane with only a few degrees of warping between the inner and outer parts of the disc. The dust however shows a larger variation in the tilt profile between small and large radii. Fig.~\ref{fig:dens_10cmthick} compares the gas and dust column density in the $xy$-plane (top) and $xz$-plane (bottom) after $\sim 900$ binary orbits. The different twist and tilt profiles between gas and dust are visible as well as the dust pile ups.

The second row of Fig.~\ref{fig:largethick} shows the same radial profiles for the St $\sim 100$ case and $H/R=$ 0.1. We note that the twist angle radial profiles show the same trend as the St = $10$ case. However, we notice that both the dip and subsequent pile in the dust surface density profile are less prominent. This is due to the fact that larger dust grains are less affected by the drag force exerted by the gas. Indeed, the third row of Fig.~\ref{fig:largethick} shows the same radial profiles for the largest grain size we simulate in this paper St $\sim 1000$, that are essentially decoupled from the gas. We can see that the effects on the dust surface density profile diminished even further.

%\begin{figure*}
%  \begin{center}
%    \resizebox{170.0mm}{!}{\mbox{\includegraphics[angle=0]{plot1mHR1}}} 
% \caption{\label{fig:1mthick}
% Radial profiles of (left to right) surface density, twist, and tilt angles for gas (solid) and dust (dashed) at 4 different snapshots for the 1 meter dust and $H/R=$ 0.1 case}
% \end{center}
%\end{figure*}

%\begin{figure*}
%  \begin{center}
%    \resizebox{170.0mm}{!}{\mbox{\includegraphics[angle=0]{plot10meterHR1}}} 
% \caption{\label{fig:10mthick}
% Radial profiles of (left to right) surface density, twist, and tilt angles for gas (solid) and dust (dashed) at 4 different snapshots for the 10 meter dust and $H/R=$ 0.1 case}
% \end{center}
%\end{figure*}

\begin{figure*}
  \begin{center}
  \resizebox{170.0mm}{!}{\mbox{\includegraphics[angle=0]{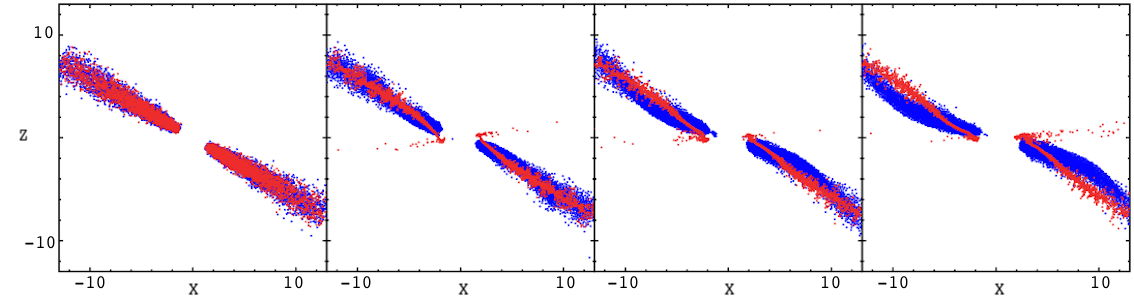}}} 
  \resizebox{170.0mm}{!}{\mbox{\includegraphics[angle=0]{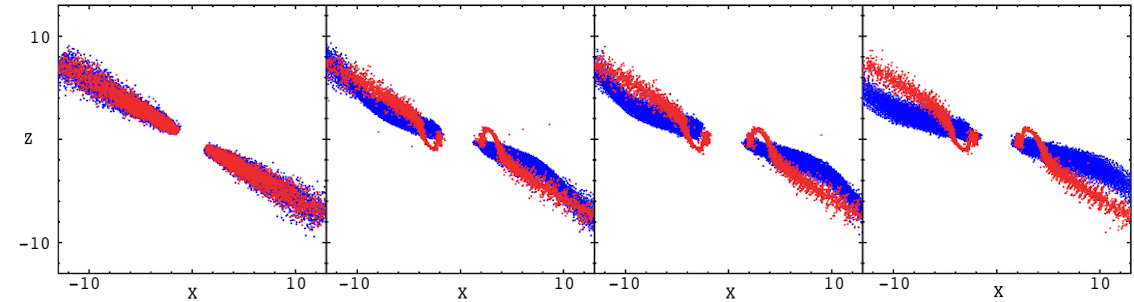}}} 
    \resizebox{170.0mm}{!}{\mbox{\includegraphics[angle=0]{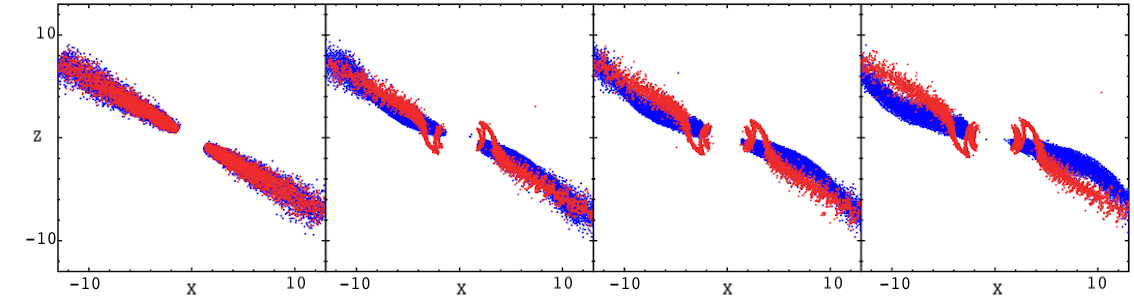}}} 
 \caption{\label{fig:oscillations}
 Particle plots coloured by type (gas and dust are blue and red; respectively) for a cross-sectional slice at 4 different snapshots (left to right, t = 0, 50, 100, 150 binary orbits) for initial $H/R=$ 0.1. Top to bottom: St $\sim 10,100,\mathrm{and} 1000$}
 \end{center}
\end{figure*}

\begin{figure}
\resizebox{85.0mm}{!}{\mbox{\includegraphics[angle=0]{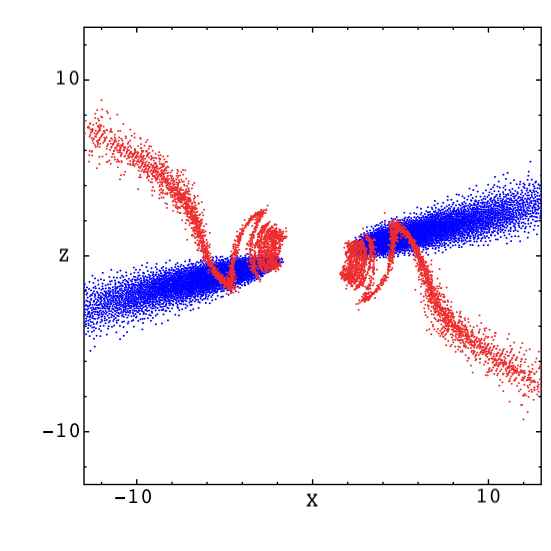}}}
\caption{\label{fig:oscillations_final}
 Particle plots coloured by type (gas and dust are blue and red; respectively) for a cross-sectional slice at the end time (after 870 binary orbits) for the St $\sim 1000$ and initial $H/R=$ 0.1 case}
\end{figure}

The second and third row in Fig.~\ref{fig:largethick} show tilt oscillations that start at the inner discs and grow and propagate with time (more evident in the St $\sim 1000$ case). In order to uncover the origin of these oscillations we show cross-sectional particle plots for 4 different snapshots starting (from the initial conditions on the left, incremented by 50 binary orbits each) for the 3 different dust species (top to bottom: $St \sim 10$, 100, and 1000) in Fig.~\ref{fig:oscillations}. Gas and dust are represented by blue and red particles, respectively. The oscillations are completely absent in the St $\sim 10$ case and increase with the grain size. We see that dust particles at the inner edge are perturbed by the binary. These perturbations grow and propagate through the disc when the dust particles are large enough not to be affected by the drag force, while they are damped quickly for grains closer to  marginal coupling. The eventual result of these perturbations is to gradually transform the dust disc to a spherical cloud above a certain grain size (final snapshot for the  St $\sim 1000$ case is shown in Fig.~\ref{fig:oscillations_final}). The dependence of the growth and extent of these oscillations on the St number as well as the initial surface density profiles is going to be investigated in a future study. 

The cross sections in Fig.~\ref{fig:oscillations} also help visualize the different precession evolution of the gas and dust discs. We can see that the fourth panel in each case (after 150 binary orbits) the gas and dust discs are already out of phase, resulting in the dips and pile ups shown in the radial profiles in Fig.~\ref{fig:largethick}. 

In the top panels we can see a few particles (a very minor fraction) that are ejected from the innermost dust ring which intersects the gas disc with a big difference in phase and tilt. The inner dust disc aligns with the binary quicker than the gas (seen in  Fig.~\ref{fig:largethick}), therefore these ejected particles keep this low inclination. We verified that these particles have high eccentricities due to this interaction. This only occurs in the case of lowest St. since the larger St dust (bottom two panels) is able able to resist this drag-induced interaction

We note that, while the inclusion of very large dust particles in a precessing disc may not be physically realistic as the dust will evolve independently from the gas before they can grow, studying the effects of large St is important for two reasons: first, depending on the gas surface density and disc size, even large St can correspond to relatively small dust particles. Second, it is not clear whether the decoupling will always happen faster than dust growth, hence including dust growth rate is of great importance in determining the largest St physically relevant in this phenomenon. We plan to investigate this in a future study. Moreover, from a theoretical point of view, it is important to study how the dynamics change for a wide range of St.

%%%%%%%%%%%%%%%%%%%%%%%%%%%%%%%%%%%%%%%%%%%%%%%%%%
\subsection{Thin discs}
\label{sec:thin}
Misaligned circumbinary gas discs with smaller aspect ratios are prone to breaking when the binary torque exceeds the viscous torque in the disc \citep{NixonEtal2013,AlyEtal2015,DoganEtal2018}. Here, we report the results of 3 simulations with the same St range as Section~\ref{sec:thick} but with $H/R=$ 0.05. In all three cases the gas disc breaks very early on, while the dust profiles are generally smoother (apart from the oscillations reported above for larger grains). 

\begin{figure}
\resizebox{85.0mm}{!}{\mbox{\includegraphics[angle=0]{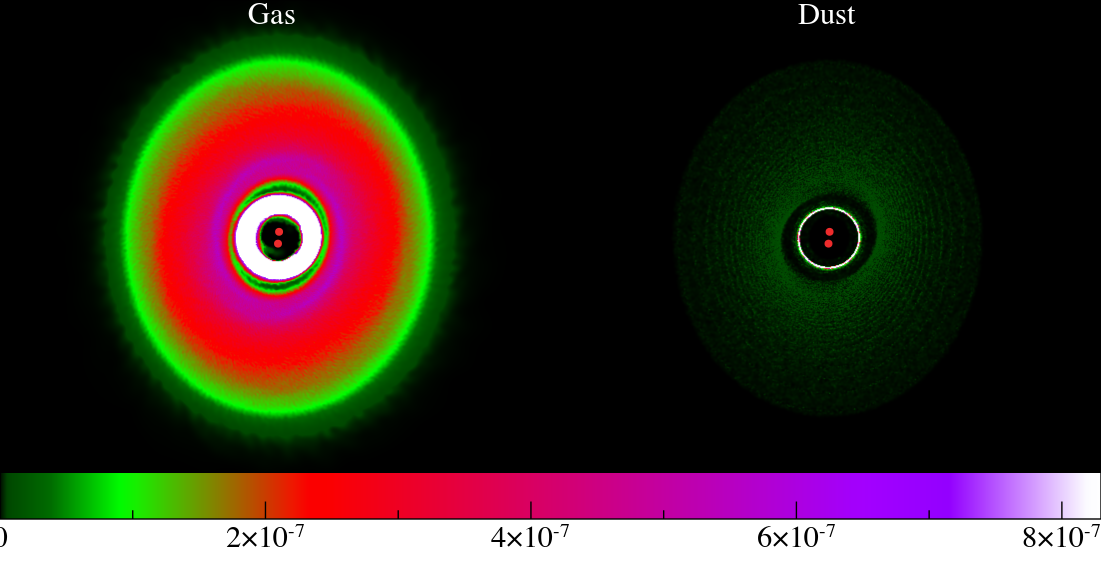}}}
\resizebox{85.0mm}{!}{\mbox{\includegraphics[angle=0]{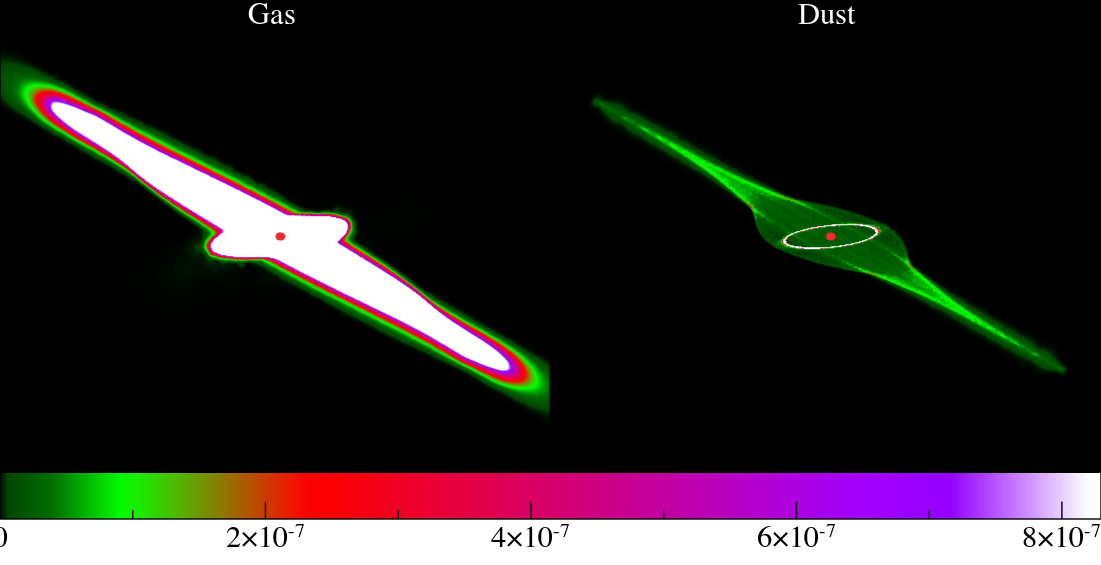}}}
\caption{\label{fig:dens_10cmthin}
  Density columns for gas (left panel) and dust (right panel) at the final snapshot for the St $\sim 10$ dust and $H/R=$ 0.05 case. Top: X-Y plane. Bottom: X-Z plane  }
\end{figure}

The immediate consequence of a break in the gas disc is the lack of viscous communication between inner and outer radii: the inner disc breaks and precesses much faster without having to `pull' the outer disc around. This is in strike contrast to the mechanism operating for thick discs reported in the previous section. Fig.~\ref{fig:dens_10cmthin} shows the gas and dust density columns for the St $\sim 10$ case (top: X-Y plane, bottom: X-Z plane). We can see that the gas disc breaks and the inner part precesses independently. The dust has a smoother profile than the gas with no sharp breaks. We note that the outer dust and gas discs remain in-phase since here the gas disc does not precess rigidly after the inner part breaks. We also see a ring of dust at higher density forming in the inner dust disc. 

%\begin{figure}
%\resizebox{85.0mm}{!}{\mbox{\includegraphics[angle=0]{splash1mHR05.eps}}}
%\caption{\label{fig:dens_1mthin}
% Density columns for gas (left panel) and dust (right panel) at the final snapshot for the 1 meter dust and $H/R=$ 0.05 case   }
%\end{figure}

\begin{figure*}
  \begin{center}
    \resizebox{170.0mm}{!}{\mbox{\includegraphics[angle=0]{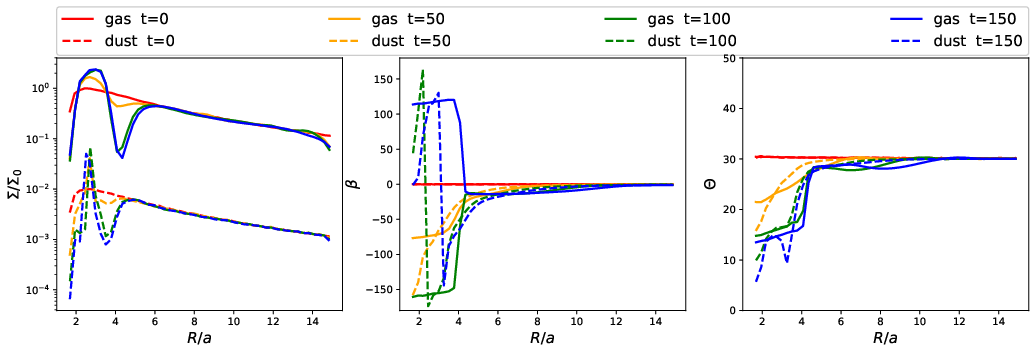}}}
    \resizebox{170.0mm}{!}{\mbox{\includegraphics[angle=0]{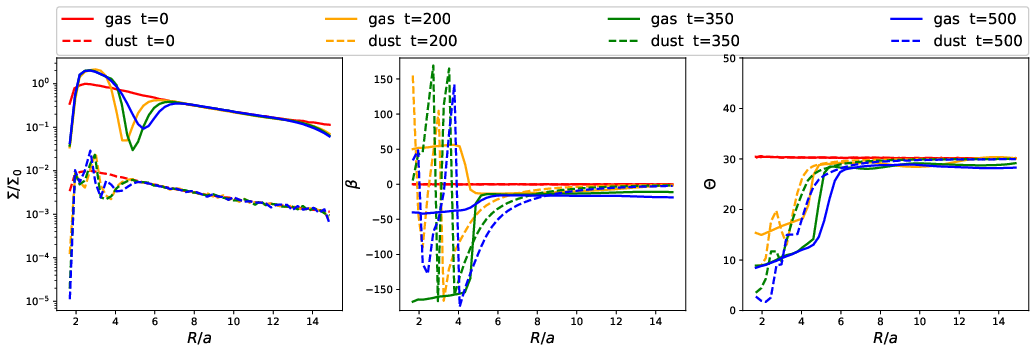}}}
    \resizebox{170.0mm}{!}{\mbox{\includegraphics[angle=0]{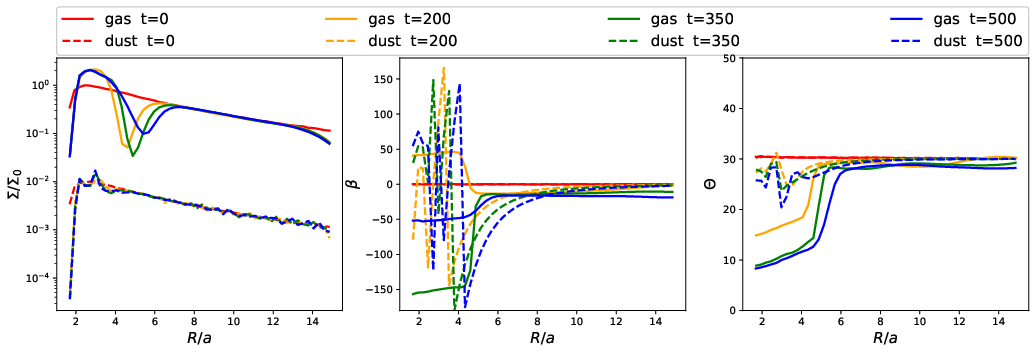}}}
 \caption{\label{fig:largethin}
 Same as Fig.~\ref{fig:largethick} but for the $H/R=$ 0.05 case. Top to bottom: St $\sim 10,100,\mathrm{and} 1000$}
 \end{center}
\end{figure*}

In Fig.~\ref{fig:largethin} we plot the radial profiles of surface density, twist and tilt angles for both dust and gas for for the three different St corresponding to the large dust grains (same as Fig.~\ref{fig:largethick} but for the thin disc). Starting from the top row (St $\sim 10$), we see that the break in the gas disc is clearly shown in the surface density profile. The inner gas disc has a very steep density gradient because of the break, which causes a very narrow dust ring at the density maximum. The twist and tilt angles profiles show a sharp break in the gas and independent precession of the inner part, whereas the dust has a smoother radial variation. We note that this particular case only ran for 150 binary orbits because the time step becomes prohibitively short as the St number drops near the break.

The second row of Fig.~\ref{fig:largethin} shows the same radial profiles but for the St $\sim 100$ case. Again we see a break in the gas disc causing steep density gradients in the inner gas disc, which forms the usual dust trap at density maxima. This effect here is less pronounced than in the St $\sim 10$ case since larger grains are less affected by the head wind and hence the dust trap is smaller. The third row indeed shows that this particle trapping is greatly reduced for the St $\sim 1000$ case, whereas the tilt oscillations in the dust are much more evident for these large dust grains. The dust in this case, being effectively uncoupled from the gas, behaves very similarly to the $H/R=0.1$ case. 

Interestingly, we note that the tearing due to binary torque only happens in the gas but not the dust component in all 3 cases. The way dust reacts to the gas break is different for each case depending on St. For the St $\sim 10$ case, the dust responds to the steep pressure gradient in the gas caused by the break by forming a dense pile up at the location of the pressure maximum, which causes two gaps in the dust at both ends of the dust trap. Note that theses gaps do not correspond to the gap in the gas component (top left in Fig.~\ref{fig:largethin}). For higher St, dust is less affected by gas drag, and hence both the dust trap and corresponding dust gaps become less pronounced, almost completely vanishing for the highest St. This can have important implications on interpreting mm and sub-mm observations. There have been reported mechanisms explaining how a gap can open in the dust but not the gas in the presence of a low-mass planet in the disc \citep{DipierroEtal2016}. Our results show that tearing due to binary torques can produce the opposite effect, a gap in the gas but not the dust.

We note that our relatively high mass discs induce some evolution in the binary evolution. At the end of all simulations the binary acquires a tilt of about 8 degrees, which explains the minimum tilt (hence maximum alignment) the disc reaches shown in all tilt profiles in this paper. The evolution in binary eccentricity and semi-major axis is however negligible. 
%\begin{figure*}
%  \begin{center}
%    \resizebox{170.0mm}{!}{\mbox{\includegraphics[angle=0]{plot1mHR05}}} 
% \caption{\label{fig:1mthin}
%  Radial profiles of (left to right) surface density, twist, and tilt angles for gas (solid) and dust (dashed) at 4 different snapshots for the 1 meter dust and $H/R=$ 0.05 case}
% \end{center}
%\end{figure*}

%\begin{figure*}
%  \begin{center}
%    \resizebox{170.0mm}{!}{\mbox{\includegraphics[angle=0]{plot10mHR05}}} 
% \caption{\label{fig:10mthin}
% Radial profiles of (left to right) surface density, twist, and tilt angles for gas (solid) and dust (dashed) at 4 different snapshots for the 10 meter dust and $H/R=$ 0.05 case}
% \end{center}
%\end{figure*}

%%%%%%%%%%%%%%%%%%%%%%%%%%%%%%%%%%%%%%%%%%%%%%%%%%
\section{Connection With Observations}
\label{sec:discuss}
%%%%%%%%%%%%%%%%%%%%%%%%%%%%%%%%%%%%%%%%%%%%%%%%%%
In section~\ref{sec:large_grains} we report two important new findings concerning weakly coupled dust particles. First, thick discs that resist breaking and precess almost rigidly in the gas experiences a different precession profile for the dust that resembles more closely that of test particles (hence a steeper radial gradient). The different precession profiles cause dust piling up where the two components are out of phase. We stress that this is a new kind of dust trap that does not correspond to a gas pressure maximum. Second, for thin discs that are prone to breaking, we show that the break only happens in the gas component while the dust piles up in the inner broken disc due to very steep gas pressure gradient. This is in line with the usual mechanism for dust traps \citep{NakagawaEtal1986}. Both phenomena are much more evident at critical St close to 1. In this section we present a simple analysis to show how to connect this critical St regime with particle size and binary separation, so that the connection with future observations is possible. In this section we envisage a general situation where the gas disc is large and does not precess but reaches a quasi-steady shape (as described in \citet{FacchiniEtal2013}). In this case the location of the warp (that does not imply disc breaking) is where the local precession time equals the sound crossing time.

A misaligned disc around a binary system experiences a radially differential precession with a frequency $\Omega_{p}$ defined as \citep{FacchiniEtal2013}:
\begin{equation}
    \Omega_{\rm p} = \frac{3\eta}{4}\frac{a^2}{R^2}\Omega_k
\end{equation}
where $\Omega_k$ is the Keplerian frequency and $\eta=M_1M_2/(M_1+M_2)^2$

For a disc with a viscosity parameter less than its aspect ratio i.e; $\alpha < H/R$, relevant for a typical protoplanetary disc, warps propagate in a wave like regime \citep{Pringle1992,Papaloizou&Lin1995,Lubow&Ogilivie2000}. The propagation speed is half the local sound speed $c_s/2$. By equating the warp propagation timescale with the precession timescale $\Omega_{\rm p}^{-1}$, we get an approximate relation for a warp radius $R_w$:
\begin{equation}
    \left(\frac{R_w}{a}\right)^{5/2-q} \sim \frac{3}{2} \frac{\eta}{\left|\frac{H}{R}\right|_a}
\end{equation}
where $q$ is the index for the sound speed power law.

\begin{figure*}
  \begin{center}
    \resizebox{170.0mm}{!}{\mbox{\includegraphics[angle=0]{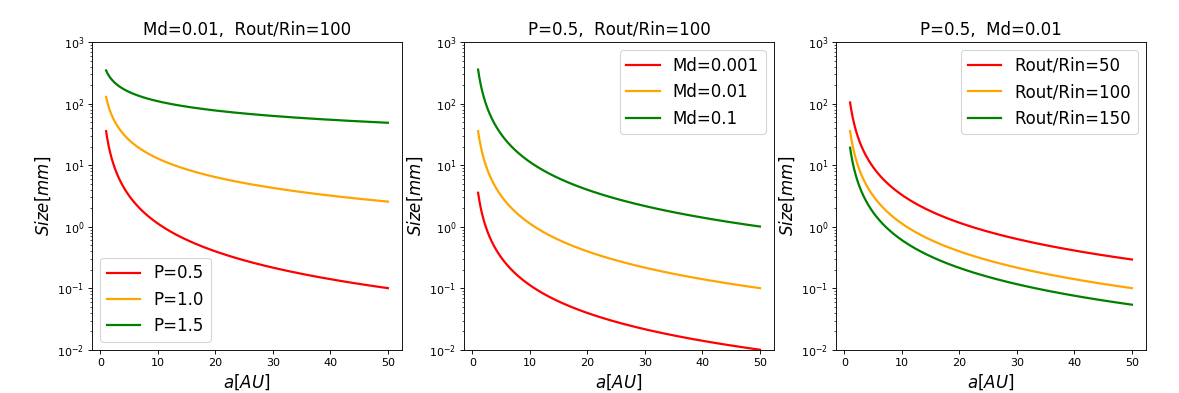}}} 
 \caption{\label{fig:radial_drift}
 Dust particle size at critical Stokes number and typical warp radii as a function of binary separation. Left to right, varying surface density power law index $p$, disc mass $M_{\mathrm{d}}$, and outer radius $R_{\mathrm{out}}$. For all models, we choose a disc aspect ratio $H/R=0.1$ at the inner radius, dust intrinsic grain density $\rho=5g/cm^2$, and sound speed power law index $q=0.25$}
 \end{center}
\end{figure*}

With this expression, and assuming a surface density profile of the form \begin{equation}
    \Sigma = \Sigma_0 \left( \frac{R}{R_{in}} \right)^{-p}
\end{equation}
we can obtain the surface density at a typical warp radius. Plugging this into the definition of Stokes number, we can estimate the dust particle size at Stokes number of order unity, i.e; where the effects of radial drift are most significant, as a function of binary and disc parameters. Fig.~\ref{fig:radial_drift} shows dust particle size at this critical St and typical warp radii as a function of binary separation for different surface density power law index $p$, disc mass $M_{\mathrm{d}}$, and outer radius $R_{\mathrm{out}}$, where we assumed a disc aspect ratio $H/R=0.1$ at the inner radius, dust intrinsic grain density $\rho=5g/cm^2$, and sound speed power law index $q=0.25$. These fixed parameters have a much lesser effect on the critical dust particle size than the 3 parameters we vary in Fig.~\ref{fig:radial_drift}.

We see from Fig.~\ref{fig:radial_drift} that the findings reported here can be observed with (sub-)mm observations (eg. ALMA) for a big range of parameters, especially for lower mass discs, less steep surface density profiles, and wider binary separations. An interesting case is the case of GG Tau A \citep{CazzolettiEtal2017,AlyEtal2018}, for which the best fit orbital parameters imply a separation of $60$ au and an almost equal mass binary (implying $\eta\simeq 0.25$). For typical disc parameters ($H/R=0.1$, $q=3/2$), this implies a warp radius of $\simeq 200$ au (coincident with the observed dust ring) and an inspection of Fig. \ref{fig:radial_drift} immediately shows that the critical dust size lies in the mm range for a wide choice of disc masses, radii and surface density slopes. 

%%%%%%%%%%%%%%%%%%%%%%%%%%%%%%%%%%%%%%%%%%%%%%%%%%
%%%%%%%%%%%%%%%%%%%%%%%%%%%%%%%%%%%%%%%%%%%%%%%%%%
\section{Small Dust Grains}
\label{sec:small_grains}
%%%%%%%%%%%%%%%%%%%%%%%%%%%%%%%%%%%%%%%%%%%%%%%%%%
\subsection{Thick discs}
\label{sec:small_thick}
\begin{figure*}
  \begin{center}
    \resizebox{170.0mm}{!}{\mbox{\includegraphics[angle=0]{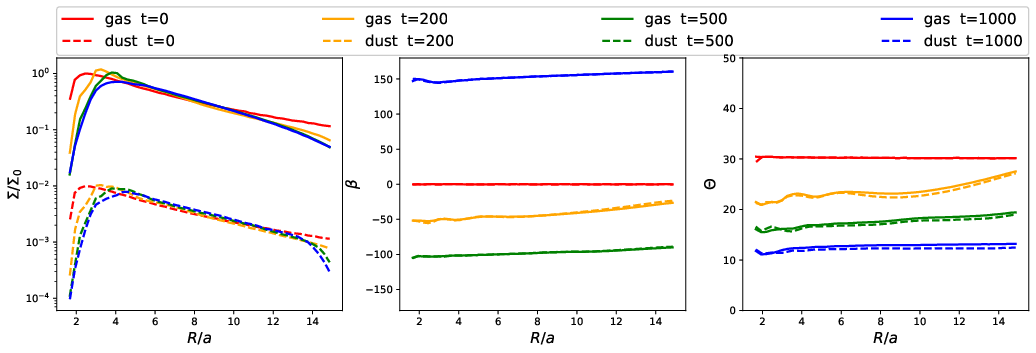}}}
    \resizebox{170.0mm}{!}{\mbox{\includegraphics[angle=0]{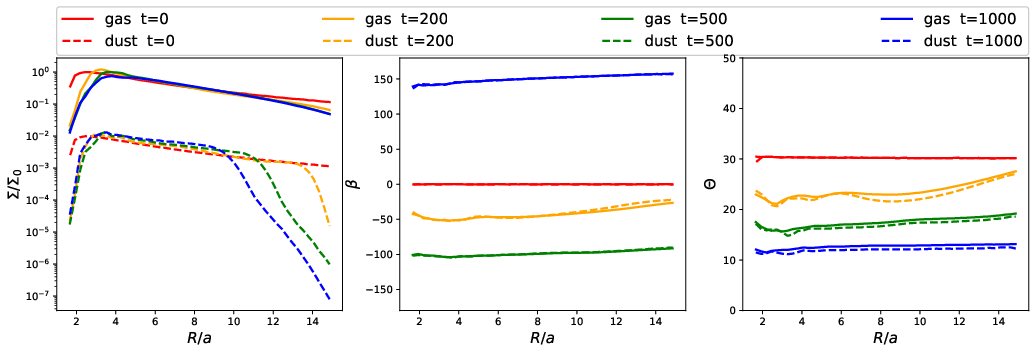}}}
 \caption{\label{fig:smallthick}
  Same as Fig.~\ref{fig:largethick}. Top to bottom: St $\sim 0.002, 0.02$}
 \end{center}
\end{figure*}

Dust species with very low St are strongly coupled to the gas because of the strong drag force they experience. We plot the radial profiles of surface density, twist, and tilt angles for the St $\sim 0.002$ case in Fig.~\ref{fig:smallthick} (top row). We can see that the dust density profile very closely matches that of the gas, with only a slight indication of a dip at the outer edge due to radial drift at the end of the simulation after 1000 binary orbits. The twist and tilt angles radial profiles are almost identical for the gas and dust as expected.

The second row in Fig.~\ref{fig:smallthick} shows the same radial profile for the St $\sim 0.02$ grains case. Here we can see that the radial drift experienced by the dust is much stronger due to the increased effect of the drag force. However, the twist and tilt angles profiles are still very similar for gas and dust, with only a slight deviation compared to the St $\sim 0.002$ case. In both cases, the disc precesses almost rigidly and smoothly goes towards alignment with the binary plane, as would be the case for a gas only disc.
%\begin{figure*}
%  \begin{center}
%    \resizebox{170.0mm}{!}{\mbox{\includegraphics[angle=0]{plot1mmHR1}}} 
% \caption{\label{fig:1mmthick}
% Radial profiles of (left to right) surface density, twist, and tilt angles for gas (solid) and dust (dashed) at 4 different snapshots for the 1 mm dust and $H/R=$ 0.1 case}
% \end{center}
%\end{figure*}
%%%%%%%%%%%%%%%%%%%%%%%%%%%%%%%%%%%%%%%%%%%%%%%%%%
\subsection{Thin discs}
\label{sec:small_thin}
\begin{figure*}
  \begin{center}
    \resizebox{170.0mm}{!}{\mbox{\includegraphics[angle=0]{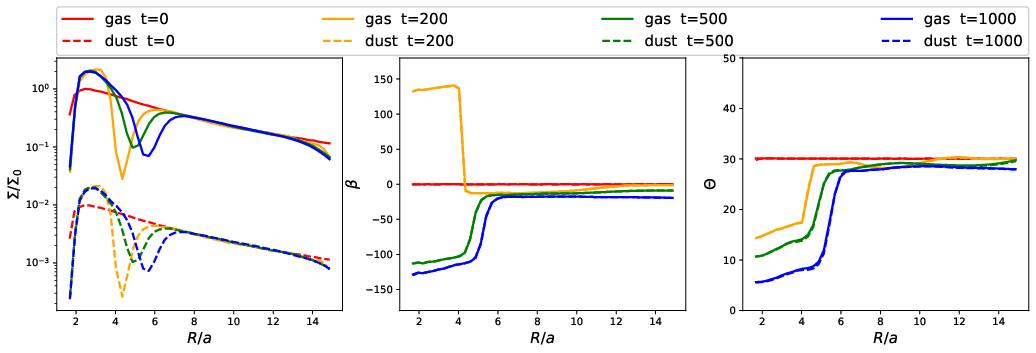}}} \resizebox{170.0mm}{!}{\mbox{\includegraphics[angle=0]{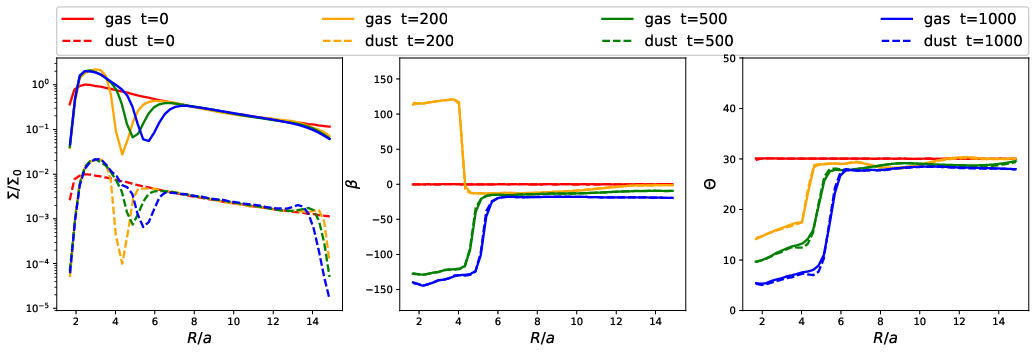}}} 
 \caption{\label{fig:smallthin}
 Same as Fig.~\ref{fig:largethick} but for the $H/R=$ 0.05 case. Top to bottom: St $\sim 0.002,0.02$}
 \end{center}
\end{figure*}

In Section~\ref{sec:thin} we showed that with the choice of parameters in this paper and an aspect ratio of 0.05 the disc breaks early on for the high St cases. We show here that low St discs also break very early on. The radial profiles for St $\sim 0.002$ (top) and St $\sim 0.02$ (bottom) cases are shown in Fig.~\ref{fig:smallthin}. In contrast to the high St simulations where the break only happens in the gas, here both gas and dust discs break at the same location with almost the same depth, due to the high coupling between gas and dust. We can also see that the St $\sim 0.02$ case shows more radial drift at the outer edge of the dust disc than the lower St case, as expected. The tilt and twist profiles are almost identical for gas and dust in both cases, although there is a slight more deviation in the tilt profile of the St $\sim 0.02$ case.
%\begin{figure*}
%  \begin{center}
%    \resizebox{170.0mm}{!}{\mbox{\includegraphics[angle=0]{plot1mmHR05}}} 
% \caption{\label{fig:1mmthin}
% Radial profiles of (left to right) surface density, twist, and tilt angles for gas (solid) and dust (dashed) at 4 different snapshots for the 1 mm dust and $H/R=$ 0.05 case}
% \end{center}
%\end{figure*}

%%%%%%%%%%%%%%%%%%%%%%%%%%%%%%%%%%%%%%%%%%%%%%%%%%
\section{Conclusion}
\label{sec:conclusion}
%%%%%%%%%%%%%%%%%%%%%%%%%%%%%%%%%%%%%%%%%%%%%%%%%%
Misaligned circumbinary discs experience differential precession due to the gravitational torque exerted by the binary. We explored the rich interplay between gas and dust in mislaligned circumbinary discs by means of hydrodynamical simulations for a range of dust sizes and 2 different disc aspect ratios. We find that for thick discs that resist breaking under the influence of gravitational torque, the gas precesses almost rigidly. Dust grains with St $\gtrsim$ 10 experience a steeper radially differential presession profile, as dust does not communicate the warp viscously. This causes enhanced radial drift when the two discs intersect and dust pile ups at the inner disc when the two discs go out of phase. We note that this is different from the usual dust trapping mechanisms that require the dust to be trapped at the gas pressure maxima. For even larger dust grains (St $\gtrsim$ 100), we notice that dust particles are perturbed close to the binary causing tilt oscillations in the inner dust disc. These oscillations grow and propagate for the largest St simulations since they are less affected by the drag force. Eventually, dust at the inner radii form a semi-spherical shape rather than a disc.

For thin discs, we find that only the gas disc breaks, but not the dust, which can have important observational signatures. The break in the gas causes a steep density gradient in the inner parts, producing a strong dust trap at the local pressure maxima. The inner gas disc precesses independently from the outer part. This means that the gas and dust discs do not precess out of phase, as was the case for thicker discs. 

In both the thick and the thin disc case, we have shown that a narrow dust ring is expected to form, although for different reasons. This may be the natural explanation for the observation of thin dust rings around binary systems, such as GG Tau A \citep{CazzolettiEtal2017,AlyEtal2018} and KH 15D \citep{Lodato&Facchini2013,Fang2019}

Highly coupled dust grains (St $\lesssim$ 0.1), on the other hand, tend to follow the gas more closely. The deviation between gas and dust tilt profiles is almost unnoticeable. However, for (St $\sim$ 0.1) we clearly see the effects of radial drift over the lifetime of the simulation.

\section*{Acknowledgments}

This project has received funding from the European Union’s Horizon 2020 research and innovation programme under the Marie Skłodowska-Curie grant agreement No 823823 (DUSTBUSTERS). The authors thanks Monash University for hospitality during the completion of this work. We thank the referee for helpful and important suggestions.

%%%%%%%%%%%%%%%%%%%%%%%%%%%%%%%%%%%%%%%%%%%%%%%%%%%%%%%%%%%%%%%%%%%%%%%%%%%%%%%%
\bibliographystyle{mnras} \bibliography{refs}
%%%%%%%%%%%%%%%%%%%%%%%%%%%%%%%%%%%%%%%%%%%%%%%%%%%%%%%%%%%%%%%%%%%%%%%%%%%%%%%%

%%%%%%%%%%%%%%%%% APPENDICES %%%%%%%%%%%%%%%%%%%%%

\appendix

% Don't change these lines

\label{lastpage}
\end{document}